\documentclass[3p,12pt,sort,compress,number,times]{elsarticle}

\usepackage{amsmath}
\usepackage[absolute]{textpos}

\DeclareMathOperator{\diag}{diag} 
\DeclareMathOperator{\trace}{Tr} 
\DeclareMathOperator{\imag}{Im} 

\begin{document} 

\begin{frontmatter}
\title{Nearest-Neighbour Interaction from an Abelian Symmetry and 
Deviations~from~Hermiticity} 

\author{G.~C.~Branco} 
\ead{gbranco@ist.utl.pt} 

\author{D.~Emmanuel-Costa} 
\ead{david.costa@ist.utl.pt} 

\author{C.~Sim\~oes} 
\ead{csimoes@cftp.ist.utl.pt}

\address{Departamento de F\'{\i}sica and 
Centro de F\'{\i}sica Te\'orica de Part\'{\i}culas (CFTP)\\ 
Instituto Superior T\'ecnico, Av. Rovisco Pais, 1049-001 Lisboa, Portugal}


\begin{abstract} 
We show that Nearest-Neighbour Interaction (NNI) textures for the quark mass
matrices can be obtained through the introduction of an Abelian flavour
symmetry. The minimal realisation requires a $\mathsf{Z_4}$ symmetry in the
context of a two Higgs doublet model. It is further shown that the NNI textures
can be in agreement with all present experimental data on quark masses and
mixings, provided one allows for deviations of Hermiticity in the quark mass
matrices at the 20\% level.
\end{abstract} 

\end{frontmatter}

\begin{textblock}{2.5}(12,1)
\small \bf
\noindent CFTP/10-002\\
arXiv:1001.5065v2
\end{textblock}

\section{Introduction} 
\label{sec:Introduction} 

In the Standard Model (SM) the flavour structure of the Yukawa couplings is not
constrained by gauge symmetry, which leads to an arbitrary flavour dependence
for fermions mass matrices, after spontaneous symmetry breaking. Finding a
framework which could explain the observed pattern of fermions masses and
mixings, is one of the fundamental open questions in particle physics. In the
last years, there has been great progress in the determination
\cite{Amsler:2008zzb} of the Cabibbo-Kobayashi-Maskawa (CKM)
matrix~\cite{Cabibbo:1963yz}. At present, essentially all data is in
agreement~\cite{Charles:2004jd} with the SM and there is clear evidence for a
complex CKM matrix, even if one allows for the presence of New
Physics~\cite{Botella:2005fc}. One may be tempted to use this input from
experiment, together with the knowledge of quark masses, to extract the flavour
structure of the quark mass matrices. The hope is to obtain, through this
bottom-up approach, a hint of a possible family symmetry constraining the
flavour structure of the SM. This bottom-up approach to the flavour puzzle is
rendered specially difficult, due to the freedom one has to make weak-basis (WB)
transformations which change the flavour structure of the quark mass matrices
$M_u$, $M_d$, while maintaining the gauge currents flavour diagonal. Needless to
say, entirely analogous considerations apply to the leptonic
sector~\cite{Branco:1999nb}. Even if there is a symmetry principle controlling
the flavour structure of the Yukawa couplings, in what WB will this family
symmetry be transparent? One of the WB which has been
proposed~\cite{Branco:1988iq} in the literature is the so-called
Nearest-Neighbour-Interaction (NNI) basis, where the elements $(1,1)$, $(1,3)$,
$(2,2)$ and $(3,1)$ vanish both in $M_u$ and $M_d$. It has been
shown~\cite{Branco:1988iq}  that in the SM, starting from arbitrary quark mass
matrices, one can always make WB transformations so that $M_u$, $M_d$ acquire
the NNI form. 

The NNI basis has been extensively studied in the
literature~\cite{Harayama:1996am} and is closely connected to the Fritzsch
ansatz~\cite{Fritzsch:1977vd} which assumes the NNI structure, together with
Hermiticity for both $M_u$ and $M_d$. Taken separately, these two assumptions do
not have any physical consequences since they are just a choice of WB. But taken
together, they do have physical implications and in fact the Fritzsch ansatz has
been ruled out by the large value of the top quark mass and the experimental
value of $V_{cb}$. In the present paper, we address the
following two questions:
\begin{description} 
\item[i)] In a multi-Higgs extension of the SM, what is the minimal scenario to
obtain the NNI structure for $M_u$ and $M_d$, as a result of an Abelian family
symmetry? 
\item[ii)] Assuming that $M_u$ and $M_d$ are in the NNI basis, what are the
minimal deviations from Hermiticity in $M_u$, $M_d$ which are required in order
to accommodate the presently available data on quark masses and the CKM matrix,
including the experimental values of $B_d-\bar{B}_d$, $B_s-\bar{B}_s$ mixings
and the measurement of the rephasing invariant phases $\beta$ and $\gamma$ of
the unitarity triangle?
\end{description} 

This paper is organised as follows. In section~\ref{sec:Symmetries}, we show
that $\mathsf{Z_4}$ is the minimal family symmetry which leads to the NNI
structure. The implementation of this symmetry requires the introduction of a
minimum of two Higgs doublets. In the section~\ref{sec:Hermiticity}, we confront
the non-zero entries of the NNI structure with experimental data. In particular,
we show that, in the NNI framework, it is possible to obtain a quark mass
spectrum and a CKM matrix consistent with all experimental data, if one allows
for relatively small deviations from Hermiticity, at the 20\% level. Finally,
our conclusions are contained in the section~\ref{sec:Conclusions}.

\section{Discrete Flavour Symmetries} 
\label{sec:Symmetries} 

In this section, we show that the NNI structure for the quark mass matrices, 
\begin{equation} 
\label{eq:nnig} 
M_u=\begin{pmatrix} 
    0 & \ast & 0\\ 
    \ast & 0 & \ast\\ 
    0 & \ast & \ast 
   \end{pmatrix}\,,\qquad 
M_d=\begin{pmatrix} 
    0 & \ast & 0\\ 
    \ast & 0 & \ast\\ 
    0 & \ast & \ast 
   \end{pmatrix}\,, 
\end{equation} 
can be achieved through the introduction of an Abelian family symmetry
in the Lagrangian. The implementations of this symmetry requires the
introduction of at least two Higgs doublets and the minimal symmetry
is $\mathsf{Z_4}$. Let $\phi_1$, $\phi_2$ denote the charges of two
Higgs doublets $\Phi_1$, $\Phi_2$, under a $\mathbf{U(1)}$ symmetry
imposed on the Lagrangian:
\begin{equation} 
\phi_1\equiv\mathcal{Q}(\Phi_1)\,,\quad \phi_2\equiv\mathcal{Q}(\Phi_2)\,. 
\end{equation} 
In order to achieve the NNI structure under this symmetry, the quark
fields must transform through the following charge assignments:
\begin{equation} 
\label{eq:phif} 
\begin{aligned} 
(q_1,q_2)&=(q_3+\phi_1-\phi_2,q_3-\phi_1+\phi_2)\,,\\ 
(u_1,\,u_2,\,u_3)&=(q_3-\phi_1+2\phi_2,q_3+\phi_1,q_3+\phi_2)\,,\\ 
(d_1,\,d_2,\,d_3)&=(q_3-2\phi_1+\phi_2,q_3-\phi_2,q_3-\phi_1)\,, 
\end{aligned} 
\end{equation} 
where $q_i\equiv\mathcal{Q}({Q_L}_i)$, $u_i\equiv\mathcal{Q}({u_R}_i)$ and
$d_i\equiv\mathcal{Q}({d_R}_i)$, with ${Q_L}_i$ denoting the left-handed quark
doublets. The charge assignments given in Eq.~\eqref{eq:phif} are not affected
by an overall change, which can be absorbed in the definition of $q_3$. In order
to preserve the zero entries of the NNI form, one must forbid
some quark bilinears to couple to the Higgs doublets. The charges
under $\mathbf{U(1)}$ of the quark bilinears ${{\overline{Q}_L}_i}_1{u_R}_j$,
${{\overline{Q}_L}_i}_1{d_R}_j$ are given for the up sector by:
\begin{equation} 
\label{eq:Gmup} 
\begin{pmatrix} 
-2\phi_1+3\phi_2 & \phi_2 &-\phi_1+2\phi_2 \\ 
\phi_2 & 2\phi_1-\phi_2 & \phi_1\\ 
-\phi_1+2\phi_2 & \phi_1 & \phi_2 
\end{pmatrix}\,, 
\end{equation} 
and for the down sector by 
\begin{equation} 
\label{eq:Gmdown} 
\begin{pmatrix} 
-3\phi_1+2\phi_2 & -\phi_1 &-2\phi_1+\phi_2 \\ 
-\phi_1 & \phi_1-2\phi_2 & -\phi_2\\ 
-2\phi_1+\phi_2 & -\phi_2 & -\phi_1 
\end{pmatrix}\,. 
\end{equation} 
It is clear from Eqs.~\eqref{eq:Gmup} and~\eqref{eq:Gmdown} that the 
non-vanishing entries in the NNI structure will be generated through the Yukawa 
couplings of the fermion bilinears with $\widetilde{\Phi}_j\equiv i 
\sigma_2\Phi^{\ast}_j$ and $\Phi_j$, for the up and down quark sectors, 
respectively. One has to further guarantee that no couplings arise for the 
zero textures of the NNI structure. It can be readily verified that the minimal 
discrete symmetry leading to the NNI structure is a $\mathsf{Z_4}$ symmetry, 
under which: 
\begin{equation} 
\Phi_j\longrightarrow\,\Phi^{\prime}_j= e^{i\:\frac{2\pi}{4}\phi_j}\,\Phi_j\,, 
\end{equation} 
with the following charges: 
\begin{equation} 
\label{eq:choice} 
(\phi_1,\phi_2)\,=\,(1,2)\,, 
\end{equation} 
leading to the following charge assignments for the 
quark fields by means of Eq.~\eqref{eq:phif}: 
\begin{equation} 
\begin{aligned} 
 &(q_1,\,q_2,\,q_3)=(2,0,3)\,,\\ 
 &(u_1,\,u_2,\,u_3)=(2,0,1)\,,\\ 
 &(d_1,\,d_2,\,d_3)=(3,1,2)\,. 
\end{aligned} 
\end{equation} 
The choice for the charges $\phi_1$, $\phi_2$ in Eq.~\eqref{eq:choice}, among
others possible charge assignments, is motivated by the possibility of embedding
this $\mathsf{Z_4}$ symmetry in the framework of a $\mathsf{SU(5)}$ Grand
Unification~\cite{Georgi:1974sy}. It is worthwhile mentioning that in grand
unified theories with extended fermionic content, the required minimal Abelian
group may be larger than $\mathsf{Z_4}$~\cite{Branco:1987tv}. It has also
been considered in the literature an example of the NNI-type fermion masses
in the context of supersymmetric flavour symmetry based on the non-Abelian
group $\mathsf{Q_6}$~\cite{Babu:2004tn}.

The most general Yukawa couplings allowed by the $\mathsf{Z_4}$
symmetry are then given by:
\begin{equation} 
\begin{split} 
-\mathcal{L}_Y=& \Gamma^1_u\,{\overline{Q}_L}\widetilde{\Phi}_1{u_R}\,+\, 
           \Gamma^2_u\,{\overline{Q}_L}\widetilde{\Phi}_2{u_R}\,\\ 
           &\,+\,\Gamma^1_d\,{\overline{Q}_L}\Phi_1{d_R}\,+\, 
           \Gamma^2_d\,{\overline{Q}_L}\Phi_2{d_R}\,+\,\text{H.c.}\,,
\end{split} 
\end{equation} 
where the Yukawa matrices $\Gamma^{1,2}_{u,d}$ have the following flavour
structure:
\begin{subequations} 
\label{eq:Y} 
\begin{equation} 
\Gamma^1_{u}= 
\begin{pmatrix} 
0 & 0 & 0\\ 
0 & 0 & b_{u}\\ 
0 & {b^{\prime}}_{u}& 0 
\end{pmatrix}\,,\quad 
\Gamma^2_{u}= 
\begin{pmatrix} 
0 & a_{u} & 0\\ 
a^{\prime}_{u} & 0 &0\\ 
0 & 0 & c_{u} 
\end{pmatrix}\,, 
\end{equation} 
\begin{equation} 
\Gamma^1_{d}= 
\begin{pmatrix} 
0 & a_{d} & 0\\ 
a^{\prime}_{d} & 0 &0\\ 
0 & 0 & c_{d} 
\end{pmatrix}\,,\quad 
\Gamma^2_{d}= 
\begin{pmatrix} 
0 & 0 & 0\\ 
0 & 0 & b_{d}\\ 
0 & {b^{\prime}}_{d}& 0 
\end{pmatrix}\,. 
\end{equation} 
\end{subequations} 
After spontaneous gauge symmetry breaking, the NNI mass matrices 
are generated through the vacuum expectation values of the Higgs doublets 
$v_1\equiv\langle\Phi_1\rangle$ and $v_2\equiv\langle\Phi_2\rangle$ leading to: 
\begin{subequations} 
\label{eq:nnigm} 
\begin{align} 
M_u&=\begin{pmatrix} 
0 & v_2\, a_{u} & 0\\ 
v_2\, a^{\prime}_{u} & 0 &v_1\, b_{u}\\[2mm] 
0 &v_1\, b^{\prime}_{u} & v_2\, c_{u} 
\end{pmatrix}\,,\\ 
M_d&=\begin{pmatrix} 
0 &v_1\, a_{d} & 0\\ 
v_1 a^{\prime}_{d} & 0 & v_2\, b_{d}\\ 
0 & v_2\, b^{\prime}_{d} &v_1\, c_{d} 
\end{pmatrix}\,. 
\end{align} 
\end{subequations} 
From Eqs.~\eqref{eq:Gmup} and~\eqref{eq:Gmdown} it can be seen that
some higher dimension operators allowed by the SM gauge group and the
$\mathsf{Z_4}$ symmetry can contribute to the vanishing elements on
the quark mass matrices in Eqs.~\eqref{eq:nnigm}. One example is the
following sixth dimensional operator,
\begin{equation} 
\frac{\lambda}{\Lambda^2}\,\,{\overline{Q}_{2\,L}}\tilde{\Phi}_1u_{2\,R} 
\Phi_1^{\dagger} \Phi_2\,, 
\end{equation} 
which contributes to the (2,2) entry of $M_u$. Moreover, one can show
by working out the matrices given in Eqs.~\eqref{eq:Gmup}
and~\eqref{eq:Gmdown} that neither $\mathsf{Z_2}$ nor $\mathsf{Z_3}$
can be invoked in order to have the NNI pattern as the result of a
discrete symmetry imposed on the Lagrangian. This can been clearly
seen by considering the (1,1)-element of the bilinear matrix given in
Eq.~\eqref{eq:Gmup} and noting that for the $\mathsf{Z_2}$ symmetry
one has
\begin{equation} 
-2\phi_1+3\phi_2\,=\,\phi_2\!\pmod{2}\,, 
\end{equation} 
which allows the Higgs doublet $\widetilde{\Phi}_2$ to couple to the
bilinear ${{\overline{Q}_L}_1}{u_R}_1$. In the case of $\mathsf{Z_3}$
symmetry one gets for the (1,1)-element from Eq.~\eqref{eq:Gmup}:
\begin{equation} 
-2\phi_1+3\phi_2\,=\,\phi_1\!\pmod{3}\,, 
\end{equation} 
which then allows the doublet $\widetilde{\Phi}_1$ to couple to the bilinear 
${{\overline{Q}_L}_1}{u_R}_1$. This implies that $\mathsf{Z_2}$
and $\mathsf{Z_3}$ are excluded. 

At this point, the following comment is in order. As we have
emphasised, in the SM, the NNI structure for $M_u$, $M_d$, is just a
choice of weak basis. On the other hand, we have shown that the NNI
form for $M_u$, $M_d$, can arise as the result of a $\mathsf{Z_4}$
symmetry, in the context of a two Higgs doublet extension of the
SM. Note however, that in a general two Higgs doublet model, the form
of the Yukawa couplings $\Gamma_{u,d}^{1,2}\,$, given in
Eqs.~\eqref{eq:Y}, is not just a choice of WB, they do imply
restrictions on the scalar couplings to quarks. 

The requirement of renormalisability implies that $\mathsf{Z_4}$ has
to be imposed on the full Lagrangian, in particular on the Higgs
potential. The most general renormalisable scalar potential consistent
with $\mathsf{Z_4}$ and gauge symmetry can be written:
\begin{equation} 
\begin{split} 
\mathbf{V}=&\,\mu_1|\Phi_1|^2\,+\,\mu_2|\Phi_2|^2\,+\,\lambda_1|\Phi_1|^4\,+\, 
\lambda_2|\Phi_2|^4\\ 
&\,+\,\lambda_3\,|\Phi_1|^2\,|\Phi_2|^2\,+\, 
\lambda_4\,\Phi_1^{\dagger}\,\Phi_2\,\Phi_2^{\dagger}\,\Phi_1\,. 
\end{split} 
\end{equation} 
It is clear that the potential has acquired a new accidental global
symmetry which, upon spontaneous symmetry breaking leads to a massless
neutral scalar, at tree level. This can be avoided by soft-breaking of the
$\mathsf{Z_4}$ symmetry through the introduction of a term like
\begin{equation} 
\label{eq:soft} 
\mathbf{V}^{\prime}=\mu_{12}\,\,\Phi_1^{\dagger}\,\Phi_2\,+\text{H.c.}
\end{equation} 
Alternatively, one may introduce a singlet Higgs field which
transforms non-trivially under $\mathsf{Z_4}$.

It can be readily verified that in this model with two Higgs doublets
and a $\mathsf{Z_4}$ symmetry, it is not possible to achieve
spontaneous CP violation even if one allows for the $\mathsf{Z_4}$
soft-breaking term of Eq.~\eqref{eq:soft}. This is essentially due to
the absence of terms like
$(\Phi_1^{\dagger}\,\Phi_2\,\Phi_1^{\dagger}\,\Phi_2) \,+\,$H.c, which
are forbidden by $\mathsf{Z_4}$. Denoting the scalar vacuum by
\begin{equation} 
\langle\Phi_1\rangle= 
\begin{pmatrix} 
0\\ 
v_1 
\end{pmatrix}\,, 
\qquad 
\langle\Phi_2\rangle= 
\begin{pmatrix} 
0\\ 
v_2\,e^{i\,\theta} 
\end{pmatrix}\,,
\end{equation} 
one verifies easily that there are two minima of the potential, both CP
conserving, corresponding to $\theta=0$ or $\theta=\pi$, for $\mu_{12}<0$ or
$\mu_{12}>0$, respectively. Therefore, in this model, CP violation arises from
complex Yukawa couplings, leading to the Kobayashi-Maskawa
mechanism, see KM in Ref.~\cite{Cabibbo:1963yz}.

\section{Minimal Deviation from Hermiticity} 
\label{sec:Hermiticity} 

In this section, we investigate what are the minimal deviations of Hermiticity 
in $M_u$, $M_d$, written in the NNI form, in order to accommodate both our 
present knowledge of the CKM matrix from experiment and the value of quark 
masses. In the NNI basis, the quark mass matrices can be written as: 
\begin{equation} 
\label{eq:nni} 
M_{u}=\begin{pmatrix} 
  0 & A_u & 0\\ 
  A^{\prime}_u & 0 & B_u\\ 
   0 & B^{\prime}_u & C_u 
 \end{pmatrix}\,, 
\quad 
M_{d}=K\begin{pmatrix} 
  0 & A_d & 0\\ 
  A^{\prime}_d & 0 & B_d\\ 
  0 & B^{\prime}_d & C_d 
 \end{pmatrix} 
\,, 
\end{equation} 
where $(A,A^{\prime},B,B^{\prime},C)_{u,d}$ are all real and the matrix $K$ can
be parametrised as
\begin{equation} 
\label{eq:K}   
K=\diag\left(e^{i\kappa_1},e^{i\kappa_2},1\right)\,, 
\end{equation} 
without loss of generality. In order to parametrise deviations from
Hermiticity in the up and down sectors, we introduce the parameters: 
\begin{equation} 
\epsilon^{u,d}_a\equiv 
\frac{A^{\prime}_{u,d}-A_{u,d}}{A^{\prime}_{u,d}+A_{u,d}} 
\,,\quad 
\epsilon^{u,d}_b\equiv 
\frac{B^{\prime}_{u,d}-B_{u,d}}{B^{\prime}_{u,d}+B_{u,d}}\,. 
\end{equation} 
For a given set of quark mass matrices, a measure of how close to Hermiticity 
$M_u$, $M_d$ are, is provided by the parameter $\varepsilon$, defined by 
\begin{equation} 
\label{eq:eps} 
\varepsilon\equiv\frac{\sqrt{\left(\epsilon_{a\!\!\phantom{b}}^u\right)^2 
+\left(\epsilon_{b}^u\right)^2+\left(\epsilon_{a\!\!\phantom{b}} 
^d\right)^2+ 
\left(\epsilon_{b}^d\right)^2}}{2}\,. 
\end{equation} 
Deviations from Hermiticity in the NNI framework have been
previously considered~\cite{Branco:1992ba}, at the time where our experimental
knowledge of the CKM matrix was very limited, in particular the rephasing
invariant phases $\beta$, $\gamma$ had not been measured.

In the evaluation of the quark masses and the CKM matrix it is useful to work 
with Hermitian matrices $H_u$, $H_d$ defined by: 
\begin{equation} 
\label{eq:Hdef} 
H_{u,d}\equiv M_{u,d}\,{M_{u,d}}^{\!\!\!\!\!\!\!\dagger}\,\,\,\,, 
\end{equation} 
which have the property of $(H_{u})_{12}=(H_{d})_{12}=0$, a
signature of the NNI basis. One can see from Eq.~\eqref{eq:nni} that the matrix 
$H_u$ is a real Hermitian matrix, while $H_d$ can be written in terms of a real 
Hermitian matrix $H^0_d$ and the phase matrix $K$ as 
\begin{equation} 
\label{eq:HdK} 
H_d=K\,H_d^{0}\,K^{\dagger}\,. 
\end{equation} 
From Eqs.~\eqref{eq:K},~\eqref{eq:HdK}, it follows that: 
\begin{equation} 
\kappa_1=\arg(H_{d_{13}})\,,\quad\kappa_2=\arg(H_{d_{23}})\,. 
\end{equation} 
Hence, the matrices $H_u$ and $H^0_d$ 
are diagonalized by two real orthogonal matrices, $O_u$ and $O_d$, in the 
following way: 
\begin{subequations} 
\begin{align} 
O^{T}_u\,H_u\,O_u=\diag(m^2_u,m^2_c,m^2_t)\,,\\ 
O^{T}_d\,H^0_d\,O_d=\diag(m^2_d,m^2_s,m^2_b)\,. 
\end{align} 
\end{subequations} 
Then the CKM matrix, $V$, is simply given by
\begin{equation} 
\label{eq:V} 
V=O_u^{T}\,K\,O_d\,. 
\end{equation} 
Observing the mass and mixing hierarchy of the quarks, the
matrices $O_u$ and $O_d$ can be well approximated
by~\cite{Branco:1992ba,Georgi:1979dq}: 
\begin{subequations} 
\label{eq:app} 
\begin{align} 
\left(O\right)_{12}&\approx-\sqrt{\frac{m_1}{m_2}}\left(1-\epsilon_a-\frac{m_2} 
{m_3}\epsilon_b\right)\,,\\[2mm] 
\left(O\right)_{13}&\approx\sqrt{\frac{m_1m_2^2}{m_3^3}} 
\left(1+\epsilon_b-\epsilon_a\right) \,,\\[2mm] 
\left(O\right)_{21}&\approx 
\sqrt{\frac{m_1}{m_2}}\left(1-\epsilon_a-\frac{m_1}{m_3}\epsilon_b\right) 
\,,\\[2mm] 
\left(O\right)_{23}&\approx\sqrt{\frac{m_2}{m_3}}\left(1-\epsilon_b\right) 
\,,\\[2mm] 
\left(O\right)_{31}&\approx-\sqrt{\frac{m_1}{m_3}} 
\left(1-\epsilon_a-\epsilon_b\right)\,,\\[2mm] 
\left(O\right)_{32}&\approx-\sqrt{\frac{m_2}{m_3}} 
\left(1-\epsilon_b+\frac{m_1}{m_2}\epsilon_a\right)\,, 
\end{align} 
\end{subequations} 
where $\epsilon_a$, $\epsilon_d$ are assumed to be small and $m_i$ denote the 
quark masses, for the up and down sectors. Notice that, we have dropped for
convenience the up and down quark sector indices.

Our task is to find ``small values'' for the parameters 
$|\epsilon_{a,b}^{u,d}|$ such that the resulting CKM matrix, $V$, is in 
agreement with experiment while a correct value for the quark masses is 
obtained. The constraint of having a correct mass spectrum is easily achieved by
using the invariants of $H_u$, $H_d$, which can be readily expressed
in terms of the parameters of $M_u$, $M_d$: 
\begin{subequations} 
\label{eq:invs} 
\begin{align} 
\trace(H)\equiv&\,m^2_1+m^2_2+m^2_3\nonumber\\ 
=&2\bar{A}^2(1+\epsilon_a^2)+ 2 
\bar{B}^2 (1+\epsilon_b^2)+C^2\,,\\ 
\nonumber\\ 
\chi(H)\equiv&\,m^2_1m^2_2+m^2_1m^2_3+m^2_2m^2_3\nonumber\\ 
=&\bar{A}^4(1-\epsilon_a^2)^2+\bar{B}^4(1-\epsilon_b^2)^2 
+8\bar{A}^2\bar{B}^2 \epsilon_a\epsilon_b\\ 
&+2\bar{A}^2(1+\epsilon_a^2)\left[C^2+\bar{B} 
^2(1+\epsilon_b^2)\right]\,, 
\nonumber\\ 
\nonumber\\ 
\det(H)\equiv&\,m^2_1\,m^2_2\,m^2_3=\bar{A}^4C^2(1-\epsilon_a^2)^2\,, 
\end{align} 
\end{subequations} 
where $\bar{A}_{u,d}$ and $\bar{B}_{u,d}$ are defined as 
\begin{equation} 
\bar{A}_{u,d}\equiv\frac{A^{\prime}_{u,d}+A_{u,d}}{2}\,,\quad 
\bar{B}_{u,d}\equiv\frac{B^{\prime}_{u,d}+B_{u,d}}{2}\,. 
\end{equation} 
Once $|\epsilon_{a,b}^{u,d}|$ and quark masses are given, the quantities
$\bar{A}_{u,d}$, $\bar{B}_{u,d}$ and $C_{u,d}$ are just determined by means of
Eqs.~\eqref{eq:invs}. Thus, the matrices $M_u$, $M_d$ and consequently $H_u$,
$H_d$ are fully reconstructed with the knowledge of $\kappa_1$, $\kappa_2$ in
Eq.~\eqref{eq:K} and the CKM matrix can then be obtained from Eq.~\eqref{eq:V}.

In our numerical search, we have performed a deep scan of small values of
$|\epsilon_{a,b}^{u,d}|$ (allowing for positive and negative values), phases of
$\kappa_1$, $\kappa_2$ and quark running masses within the allowed range at
$M_Z$ scale taken from Refs.~\cite{EmmanuelCosta:2009bx,Rodrigo:1993hc}. Among
the various solutions obtained, we have accepted only those corresponding to a
correct CKM matrix. This include not only the allowed CKM moduli, but also the
experimental limits on the strength of CP violation measured by
$I\equiv\left|\imag\left(V_{us}V_{cb}V_{ub}^{\ast}V_{cs}^{\ast}
\right)\right|\,$ and on the angles $\beta \equiv
\arg(-V_{cd}V_{cb}^{\ast}V_{td}^{\ast}V_{tb})$ and $\gamma \equiv
\arg(-V_{ud}V_{ub}^{\ast}V_{cd}^{\ast}V_{cb})\,$ of the unitarity triangle.

\begin{figure}[ht]
\centering\includegraphics[width=14cm]{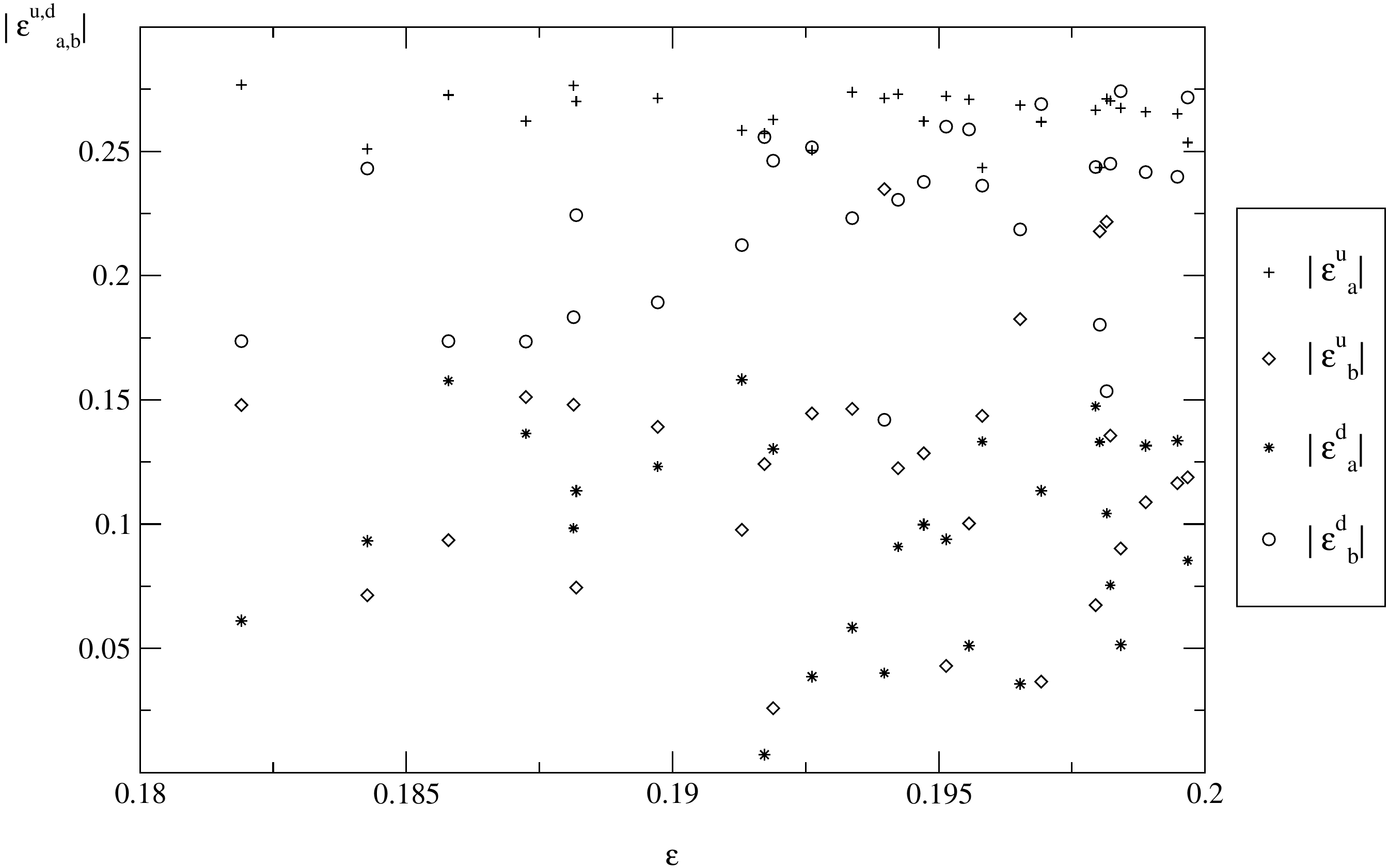}
\caption{\label{fig} Plotting a set of solutions for
$|\epsilon_{a,b}^{u,d}|$ corresponding to the constraint $\varepsilon\leq0.2$.}
\end{figure}

We have checked that most of our numerical solutions require that
$\epsilon_{a,b}^{u}$ be negative and $\epsilon_{a,b}^{d}$ be positive. Moreover,
we have also verified that the NNI basis is experimentally compatible with
having $\epsilon_{a}^{u}=0$ or $\epsilon_{b}^{u}=0$ (or even both), but this
would imply a large deviation from Hermiticity in the down sector,
$\epsilon_{a,b}^{d}\gtrsim0.6\,$. We also emphasise that no numerical solution
for $\epsilon_{a}^{d}=0$ or $\epsilon_{b}^{d}=0$ corresponding to
$\epsilon_{a,b}^{u}\leq0.3$ was found. Finally, the consistent set of
$|\epsilon_{a,b}^{u,d}|$ obtained numerically are sketched in Figure~\ref{fig},
for $\varepsilon\leq0.2\,$.

One numerical example of $|\epsilon_{a,b}^{u,d}|$ extracted from
Figure~\ref{fig} is given by,
\begin{equation} 
\label{eq:epsn} 
\begin{array}{l@{\quad}l} 
\epsilon^u_a=-0.243\,,&\epsilon^u_b=-0.144\,,\\ 
\epsilon^d_a=0.133\,,&\epsilon^d_b=0.236\,,\\ 
\end{array} 
\end{equation} 
which corresponds to $\varepsilon=0.196$ by Eq.~\eqref{eq:eps}. Using as input
the running quark masses taken at $M_Z$ scale within the allowed range:
\begin{equation}
\label{eq:massesn} 
\begin{array}{l@{\quad}l}
m_u=2.0\,\text{MeV}\,, & m_d=2.7\,\text{MeV}\,,\\
m_c=0.557\,\text{GeV}\,, & m_s=47\,\text{MeV}\,,\\
m_t=168.3\,\text{GeV}\,, & m_b=2.92\,\text{GeV}\,,
\end{array} 
\end{equation} 
together with the invariants in Eqs.~\eqref{eq:invs}, 
one obtains for ${\bar{A}}_{u,d}$, ${\bar{B}}_{u,d}$,
$C_{u,d}$:
\begin{equation} 
\label{eq:abcn} 
\begin{array}{l@{\quad}l} 
{\bar{A}}_u=34.4\,\text{MeV}\,,&
{\bar{A}}_d=11.5\,\text{MeV}\,,\\
{\bar{B}}_u=9.76\,\text{GeV}\,,&
{\bar{B}}_d=0.371\,\text{GeV}\,,\\
C_u=167.7\,\text{GeV}\,, & C_d=2.87\,\text{GeV}\,. 
\end{array} 
\end{equation} 
The matrices $H_u$, $H_d$ are then fully determined from Eqs.~\eqref{eq:epsn}, 
\eqref{eq:abcn} by taking the input phases $\kappa_1$,
$\kappa_2$, 
\begin{equation} 
\kappa_1=-121.7^{\circ}\,,\quad\kappa_2=-21.0^{\circ}\,. 
\end{equation} 
Thus, the CKM matrix, $V$, is
directly evaluated from Eq.~\eqref{eq:V}, obtaining:
\begin{equation} 
\label{eq:Vckmnum} 
\left|V\right|\,=\,\begin{pmatrix}
0.9743 & 0.2253 & 0.0034 \\ 
0.2251 & 0.9734 & 0.0415 \\ 
0.0087 & 0.0407 & 0.9991 
\end{pmatrix}\,. 
\end{equation}
For the rephasing invariant angles and the strength of CP violation, one
obtains: 
\begin{equation}
\label{eq:cpvn} 
\begin{aligned} 
\alpha&=89.7^{\circ}\,,\\
\sin(2\beta)&=0.669\,,\\ 
\gamma&=69.3^{\circ}\,,\\
I&=2.92\times10^{-5}\,.
\end{aligned}
\end{equation}

The results of Eqs.~\eqref{eq:massesn}, \eqref{eq:Vckmnum},~\eqref{eq:cpvn}
for quark masses and their mixings are in agreement with
experiment~\cite{Amsler:2008zzb}.
They were obtained by exact numerical diagonalisation of the quark mass
matrices, with no approximations involved. Yet, it is instructive to understand
the reason why one can reproduce in the NNI framework a correct CKM matrix, $V$,
with relatively small deviations of Hermiticity. Note that if one assumes exact
Hermiticity in the NNI basis, one is led to the Fritzsch Ansatz~(FA) which has
been ruled out by experiment. As previously mentioned, the main reason why FA
has been excluded, has to do with the experimental value of $|V_{cb}|$ and the
fact that the top quark is very heavy. In the framework of the NNI with small
deviations from Hermiticity, using Eqs.~\eqref{eq:V} and~\eqref{eq:app}, one
obtains:
\begin{equation} 
\begin{aligned} 
|V_{cb}|\cong\,\left|O^u_{22}O^d_{23}e^{i\kappa_2}+O^u_ {32}O^d_ 
{33}\right|\,. 
\end{aligned} 
\end{equation} 
Taking into account that $O^d_{23}\approx\sqrt{m_s/m_b}(1-\epsilon^d_b)$ and 
$O^u_{32}\approx-\sqrt{m_c/m_t}(1-\epsilon^u_b)$, it is clear that in order to 
obtain $|V_{cb}|$ consistent with experiment, one needs to obtain a 
``suppression'' of the $O^d_{23}$ and an enhancement of $O^u_{32}$. This is 
achieved for the values of $\epsilon^{u,d}_{a,b}$ given in Eq.~\eqref{eq:epsn}. 
 
\section{Conclusions} 
\label{sec:Conclusions} 

We have pointed out that the NNI form for the quark mass matrices can
be obtained in the context of a two Higgs doublet extension of the SM,
through the introduction of a $\mathsf{Z_4}$ symmetry. We have further
shown that the NNI scheme, with small deviations from Hermiticity, can
correctly reproduce the experimentally allowed values for quark masses
and CKM mixings.

Most of the searches for allowed fermion mass textures have been
conducted in the framework of Hermitian or symmetric quark mass
matrices. In the SM, quark mass matrices need not be Hermitian or
symmetric. But in the framework of a left-right symmetric
theory~\cite{Pati:1974yy} or $\mathsf{SO(10)}$ Grand
Unification~\cite{Georgi:1975qb}, Hermitian or symmetric quark
matrices naturally arise. Non-Hermitian quark mass matrices have also
been considered and their phenomenological implications analysed in the
literature~\cite{Branco:1994jx}. We find remarkable that a good fit of
quark masses and mixing is obtained in the NNI framework, with small
deviations of Hermiticity, specially taken into account the rather
precise experimental information one has at present on the CKM matrix,
including those resulting from the measurements of $B_d-\bar{B}_d$,
$B_s-\bar{B}_s$ mixings and the rephasing invariant phases $\beta$ and
$\gamma$.

\section*{Acknowledgements}
This work was partially supported by Funda\c c\~ao para a Ci\^encia e a 
Tecnologia (Portugal) through the projects CERN/FP/83503/2008 and CFTP-FCT 
Unit 777 which are partially funded through POCTI (FEDER), by Marie Curie RTN 
MRTNCT-2006-035505.

\end{document}